Markov chain models for inspecting response dynamics in psychological testing.

Andrea Bosco

Dipartimento di Scienze della Formazione, Psicologia, Comunicazione

Università degli Studi di Bari – Aldo Moro

Abstract

The importance of considering contextual probabilities in shaping response patterns within psychological testing is underscored, despite the ubiquitous nature of order effects discussed extensively in methodological literature. Drawing from concepts such as path-dependency, first-order autocorrelation, state-dependency, and hysteresis, the present study is an attempt to address how earlier responses serve as an anchor for subsequent answers in tests, surveys, and questionnaires. Introducing the notion of non-commuting observables derived from quantum physics, I highlight their role in characterizing psychological processes and the impact of measurement instruments on participants' responses. We advocate for the utilization of first-order Markov chain modeling to capture and forecast sequential dependencies in survey and test responses. The employment of the first-order Markov chain model lies in individuals' propensity to exhibit partial focus to preceding responses, with recent items most likely exerting a substantial influence on subsequent response selection. This study contributes to advancing our understanding of the dynamics inherent in sequential data within psychological research and provides a methodological framework for conducting longitudinal analyses of response patterns of test and questionnaire.

1. Introduction

Tests and questionnaires are instruments widely used in a very large number of health and social contexts. They are commonly based on an ordered sequence of standardized questions or statements designed to elicit responses that are compatible with certain characteristics of the individual being tested. Test items should prompt a response that aligns with an approximate representation of the individual's condition in relation to the characteristic of interest. Tests are typically designed to minimize the influence of extraneous variables that could anomalously affect individuals' responses and, consequently, limit the reliability and validity of the measurements. Randomizing the order of questions is one of the strategies used to o minimize the possibility of cheating and enhance test. Nonetheless, the positioning of the items within a test may have an impact on the responses being a source of systematic measurement error (e.g., Ackerman et al., 2013; Streiner et al., 2016). To guide responses, clear and consistent instructions can be provided, and control scales can be introduced, such as those assessing the tendency of confirmation bias (Paulhus, 1991), and sophisticated score standardization procedures can be used to reduce error variability in measurement, as in the case of applying Item Response Theory models. However, it is important to note that a certain degree of variability in responses is expected. For instance, psychological states can influence traits and fluctuate to some extent over time. Test developers take all of this into account by focusing on the overall pattern of responses, which, in practice, involves the use of the sum scale, rather than individual responses to specific questions or items. Despite all the efforts made, this variability is quite extensive, and it is typically far from being explained completely by the amount of shared variance among items associated to the underlying latent factor.

Searching for new sources of variance: The dynamics of Test Responses

In the social science literature, it is widely accepted that a questionnaire - and by extension, a test - represents an environment in which the respondent must navigate to provide the answers the researcher solicits (Schwarz & Sudman, 2012). In this dynamic process of questions and answers, there may be a somewhat coherence that is only partially explained by the content of the questions themselves. In fact, the researcher can take in to account the response process, namely, how past experiences and decisions can influence locally the responses.

In this perspective we can consider path dependency regards how past experiences can shape future outcomes. Choices made in the past limit future choices, creating a "path" that becomes increasingly difficult to deviate from. Extending this claim, shifting from the context of social studies to legal sciences, there are studies showing path-dependency in verdicts by a jury in unrelated cases (Bindler & Hjalmarsson, 2019). The authors argue that there is a positive autocorrelation between two consecutive independent verdicts. This positive dependence can be explained, according to the authors, by the jury's desire for internal consistency or the emotional impact of one case on another. Moreover, when we analyze sequential data, autocorrelation can be used. It refers to the degree of correlation of a time series variable with itself, measuring how the lagged version (by one or more steps) of the observed values of a variable is related to the original version. Du and Clark (2017) refer to the concept of autocorrelation, detecting in a study of implicit learning that an interpretative model based on autocorrelation is more effective than models based on widely recognized concept of chunking. They observed significant first-order autocorrelation effects in their study, which contribute to explaining how internal cognitive representations tend to organize knowledge through first-order contingencies in the presence of sequences. Consistently, the instantaneous value of a quantity determined by another variable depends on its value at the same moment and on the values manifested in previous steps (and particularly to the last one) is called hysteresis (lag), a concept originating from physics but adopted by other disciplines (e.g., economics). Thiel and collaborators (2014) talk about hysteresis, adopting a variant of the classic two-point discrimination test and attributing part of the way participants solved the task to hysteresis. They were more likely to respond in the same way between two subsequently presented trials. The authors clarify that, on average, stimulus properties explained a significantly larger part of variance than hysteresis. However, when they observed participants' behavior in trials where the distance between the two points was close to the discriminability threshold, hysteresis explained a larger share of variance compared to that explained by stimulus properties. Thus, hysteresis would act as an implicit antecedent or a primer in the decision-making process in this tactile perceptual task, becoming central when the explicit stimulus properties appear inconclusive. Hysteresis also emerges in a study of dynamic psychophysics on numerical discrimination proposed to five-year-old children by Odic and collaborators (2014). According to the authors, numerical discrimination capacity is state-dependent, relying on previous resolution experiences. A sequence of incorrect decisions worsens performance regardless of the current trial's difficulty, while prolonged experience of correct decisions improves it.

In conclusion, the concepts of path dependency, first-order autocorrelation, state-dependency, and hysteresis all converge in highlighting the influence of previous experiences in attempting to formulate a response to a simple (perceptual discrimination tasks or implicit learning) or complex questions (making a verdict in a forensic context, answering surveys or tests). In the ensuing paragraph, I aim to expound furtherly on how responses provided earlier serve as the foundation for subsequent answers in tests, surveys, and questionnaires. In a 2012 paper, Atmanspacher and Römer propose an explanatory model of how humans would treat the order in sequential information, especially concerning properties or observables that do not commute. The authors define non-commuting observables as those properties of the measured systems – in psychology, mainly persons – that are significantly altered by the measuring instrument. They borrow this concept from quantum physics, where measurement processes can produce commuting or non-commuting observables, whereas classical physics only allows measurement instruments that do not modify the measured system. According to this view, all psychological processes measured with surveys or tests should generate non-commuting observables. Let's try an example: if we were to measure a person's body temperature using a mercury thermometer, we might consider the influence of the measuring instrument on the system to be weak and consequently negligible. However, if we ask questions to the same person who must organize the information and provide an answer, we must assume that the instrument's influence is strong, and therefore, the measurement itself generates a significant modification in the system. This is why the authors conclude that almost all areas of social and psychological research are characterized by non-commuting observables. The authors argue that this concept should already be widely accepted, given that an extensive part of the methodological literature has focused on the topic of order effects since the mid-20th century (e.g., Brainerd, 1979; Hogarth & Einhorn, 1992). An order effect occurs when a participant's responses to a certain experimental condition are influenced by the previous conditions they have been exposed to. By extension, it is assumed here that this happens not only between conditions but also, to some extent, between successive trials. Atmanspacher and Römer refer to these circumstances as contextual probabilities to emphasize the role that previous experiences play an important role in influencing the probability of a certain response manifesting in the system. Although quantum theory has also been invoked to try to define a theoretical framework for context effects in responses to ordered sequences of stimuli, to the best of our knowledge, there are no studies that use statistics able to capture the portion of information derived from questionnaires and tests that can be essentially attributed to the sequence of responses itself. In other words, capturing the dynamics of survey and test responses. Sequences of information provided by a series of statistical units can be analyzed by time series models. Time series analysis involves studying the patterns, trends, and dependencies within a dataset over time. One of the most employed approaches is Autoregressive Integrated Moving Average (ARIMA) modeling (e.g. Shumway, et al. 2017). Markov chain models are also used for analyzing sequences of information, particularly in the context of stochastic processes where the probability of transitioning from one state to another depends solely on the current state (Markov property). These models can be valuable for capturing and predicting the dynamics of sequential data over time, making them suitable for certain types of longitudinal analyses. This kind of model is known as first-order Markov Chain. Higher-order Markov Chain are also available: the probability of transitioning to a future state depends on multiple preceding states, providing a more nuanced representation of sequential dependencies.

In this study, we will discuss the first-order Markov chain model used to describe a sequence of events where the probability of the next event depends only on the current state of the system, and not on the history of preceding events. This choice is warranted by the circumstance that an individual's recollection, while responding to a test or questionnaire, exhibits partial attentiveness to the provided responses and the most near preceding item is prone to exert a more pronounced influence on the subsequent response selection (Tourangeau, Couper, & Conrad, 2004).

Markov chains for the analysis of the dynamics of Test Responses

A Markov chain is a stochastic process, in our case with discrete states, that satisfies the Markov property, meaning that the probability of transitioning from one state to the next depends only on previous state and not on all the previous states.

In my view, the states could be represented by response scale categories. Often, these are ordinal, or Likert-like interval scales used to measure the degree of agreement with certain test items, which are presented in an ordered and sequential manner.

The use of Markov chains for the analysis of data in a social or psychological study perspective is not new (e.g., Jiang, 2009; Kintsch, & Morris, 1965). Nonetheless, to the best of our knowledge, this is the first time that a Markov Chain is proposed from the perspective of the study of response dynamics and applied to the psychological testing. The Markov chain is expected to capture fluctuations independent of the relationship between items and the underlying construct. The process that generates the passage from one state to another (or to remain in the same state) is called transition. Transition probabilities are summarized in an adjacent matrix r×c, where r=c. The transition matrix has the following properties:

Conditional probabilities

The sum of each row is equal to 1. This is due to the conditional nature of transition probabilities. Since the rows represent each starting state and the columns represent all possible destination states.

Symmetry and reversibility

The matrix is symmetric if the process is reversible. If the matrix is symmetric, it means that the transition probability from one state to another is equal to the reverse transition probability. In the specific case of test responses, reversibility from one state to another cannot be excluded in principle. Therefore, in our case, we can say that the matrix may be symmetric even though it is not a necessary condition.

Diagonal dominance and ergodicity

The matrix is diagonally dominant if the process is ergodic. A matrix is diagonally dominant if the probabilities on the diagonal are non-zero. The probabilities on the diagonal express the probability of remaining in the same state in two successive phases. We define this occurrence self-transitions as inertia, which is an aspect that may interest researchers in the analysis of test response dynamics. A process is ergodic if, regardless of its initial condition, its final probability distribution is independent from time or simply by repetitions. This means that it is possible to observe, through appropriate transformations, whether the transition matrix stabilizes over time: regardless of the starting state, a certain destination state can be observed. In practical terms, this would be a vector with as many cells as the finite number of states. This probability distribution is called stationary or in an equilibrium state. There is thus an intimate relationship between diagonal dominance and ergodicity of the process. Diagonal dominance ensures that there are non-zero probabilities of remaining in the same state, while ergodicity implies that the chain has a single communicating class (the transition from any state to any other is allowed) and is aperiodic, which leads to the existence of a stationary distribution. These concepts are important for understanding long-term behavior and convergence properties of Markov chains. The study of the stationary distribution can also be of interest in the analysis of test response dynamics. Finding the stationary distribution involves solving a set of linear equations or to adopt specialized numerical techniques and algorithms, such as the eigenvalue decomposition method (e.g., Madras, & Randall, 2002), that can converge to the stationary distribution efficiently for certain types of Markov chains. Alternatively, a very simple method can be implemented to obtain stationary distribution, that is, the power iteration method (Nesterov & Nemirovski, 2015).

Transition matrices offer mathematical tools to address questions often overlooked in traditional psychological test theories. For example: 1) Compare transition matrices across populations, assessing differences in: Inertia that compare probabilities on the diagonal of transition matrices to reveal differences in response consistency across different groups; and Stationary distributions functional to understand how response dynamics differ between populations in the long run. 2) Identify population origin based on response dynamics, calculating the likelihood that an individual's response sequence suggests they belong to a particular population. In such cases, The Markov chains constitute a classical yet innovative solution within the realm of studies on the dynamics of test and questionnaire responses. Moreover, the same technique can have wide application in health studies approaching differential diagnosis assessment when, for example, the test content appears to be poorly or barely discriminative. 3) Alternatively, a more theoretical approach can be used without requiring observed data for training: a) Theoretical model comparison that directly compare theoretical models of response dynamics against models where the data exhibits no clear structure. This can be determined when all state transitions are equally likely, resulting in maximal entropy within the transition probability matrix; b) Simulating test responses of a specific length according to a defined stochastic discrete probability distribution. Considering the premises, this study aims to explore a couple of potential applications of Markov chain models in analyzing response sequences from a self-report test.

2. Different response dynamics in two populations. The case of the Four Factors Mind Wandering Questionnaire

The Four Factors of Mind Wandering Questionnaire (4FMW, Lopez et al., 2023) was administered together with a battery of other psychological instruments to three cohorts of undergraduate psychology students with the aim of promoting self-awareness of possible mental disorders potentially hindering their academic career. MW is a phenomenon that occurs when the mind disengages from the current activity, instead focusing on thoughts, memories, or fantasies that are not related to the task itself. The effects of this phenomenon can be positive, such as in the case of creative daydreaming (McMillan, Kaufman, & Singer, 2013) or it can have negative effects such as promoting distraction, anxiety, and depression, decreasing control over the present state and self-awareness, promoting also obsessive thoughts. The 4FMW intercepts four facets of the maladaptive MW, that is, regarding poor attentional control: a) unawareness, b) inattention, and regarding (guilty) fear of failure: c) with objects and activities, d) in personal relationships. All together, they provide an assessment of general MW. Therefore, MW has also recently been considered both a sign of obsessive-compulsive symptomatology (Cole, & Tubbs, 2022), as well as of attention deficit hyperactivity disorder (ADHD) symptomatology (Osborne et al., 2023). It is possible that people with probable OCD and probable ADHD differ based on the test response dynamics rather than in the total score on MW test. For example, the former may show higher levels of inertia than the latter and a higher degree of coherence in the response sequences, so that there is therefore a higher level of hysteresis in one of the two populations, that is, the students with the signs of a probable OCD.

2.1. Materials and Methods

The accomplished assessment procedure provided indications also for two groups of students with signs of probable OCD or of probable ADHD - among others probable diagnosis. Data of nine students showing a profile compatible with both the disorders are disregarded. Both groups showed significantly higher levels in 4FMW compared to the control groups but slightly different from each other. This is a case in which studying the dynamics of responses can be opportune. One-hundred students (80 women, the gender imbalance is due to the original composition of the cohorts prevalently constituted by women) showed signs of possible obsessive-compulsive disorder and of probable Attention-deficit/hyperactivity disorder (prob. OCD n = 27, prob. ADHD n = 73; these proportions are compatible with the lifetime prevalence in the Italian general population: 2-3% for OCD and 8-9% for ADHD taking into account the general sample size, approximately 650 people).

The 4FMW questionnaire (Lopez et al., 2023) is composed of 16 items regarding four content areas, two of which associated with poor attentional control (unawareness, inattention) and two with guilty fear of failure (Failure in social interaction and in the use of object and daily life activities). Poor attentional control and guilty fear of failure are considered the negative facets of MW's McMillan, Kaufman and Singer (2013) model.

Examples of items are: "Do not remember part of a conversation you were following, realizing that you were not paying attention", "Put back an object in the wrong place (e.g.; put the keys in the wardrobe)", "Are not aware of what is happening around you", "Are not able to focus your attention on what you're reading, and to have to read again". the fact that the consecutive items in a test come from different content areas does not mean that precludes the possibility of observing hysteresis, since it should be, at least partially, independent from the content of questions.

Signs of possible obsessive-compulsive disorder were assessed with DSM-5 Self-Rated Level 1 Cross-Cutting Symptom Measure – Adult (Italian version: DSM-5 Scala di valutazione dei sintomi trasversali di livello 1 autosomministrata – Adulto; Fossati, Borroni, & Del Corno, 2015) with Level 2 – Repetitive Thoughts and Behaviors – Adult (adapted by the Florida Obsessive Compulsive Inventory [FOCI] Severity Scale [Part B]; Italian version: Livello 2 – Pensieri e comportamenti ripetitivi – Adulto; Fossati, Borroni, & Del Corno, 2015). Signs of possible ADHD were assessed through the Migliarese and collaborators's Italian version of the Ultra-short screening list for ADHD in adults (2015).

The statistical approach used here is inspired by Paxinou and collaborators (2021). As in any conventional test, the sequence of items is the same for each participant, consequently, the sequence represents a constant of the measurement instrument. The space of the states was as follow:

$$S = \{never, almost\ never, sometimes, often, alwais\}$$

[1]

corresponding to the response modalities to the item. For the sake of simplicity each state is resembled in the subsequent tables with a number from 1 to 5, respectively.

$$S = \{1, 2, 3, 4, 5\}$$

[2]

The most relevant characteristic of Markov chains described earlier is represented in the following formula, known as the Chapman-Kolmogorov formula (Karush, 1961):

$$P(X_n = x_n \mid X_1 = x_1, X_2 = x_2, \ldots, X_{n-1} = x_{n-1}) = P(X_n = x_n \mid X_{n-1} = x_{n-1})$$

[3]

The formula [3] can be used to calculate the probability of any event in a stochastic process, given the transition probability between the states of the process. The transition probabilities would then represent the probability of transitioning from one state to another. Transition probabilities can be calculated using a variety of methods, depending on the specific system being modeled. In some cases, it is possible to calculate the transition probabilities directly from data.

$$P(X_n = j \mid X_{n-1} = i) = p_{ij}$$

[4]

That is, the probability P that the process transitions from state i to state j is equal to pij. This probability is conditional. Transition probabilities are summarized in an adjacent matrix i×j, where i=j. The elements of the matrix are the transition probabilities from one state to another.

$$P = \begin{bmatrix} p_{11} & \cdots & p_{1j} \\ p_{21} & \ddots & p_{2j} \\ p_{i1} & \cdots & p_{ij} \end{bmatrix}$$

[5]

The starting point is the sequence of responses given to the test items, which are always presented in the same order for all participants. For example, participant O05 of our overall data set generated the following sequence of responses: 3243232443244333. From the sequence it is possible to pass to the transition matrix for each individual participant:

Transition Probability Matrix of student O05

|  |  | State j | | | | |
|---|---|---|---|---|---|---|
|  |  | 1 | 2 | 3 | 4 | 5 |
| State i | 1 | - | - | - | - | - |
|  | 2 | - | - | 0.25 | 0.75 | - |
|  | 3 | - | 0.67 | 0.33 | - | - |
|  | 4 | - | - | 0.60 | 0.40 | - |
|  | 5 | - | - | - | - | - |

[6]

The transitions in a 16-item test are 15. Each row represents the current state, while each column represents the next state. For example, it is easy to verify that in all cases where the transition occurs from state 3 (third row of the matrix) to itself or to another state, in 2/3 of the cases it is a transition from state 3 to state 2 and in 1/3 of the cases it is the permanence in the same state. No other outcomes are observed from state 3. Since these are conditional probabilities, each row of the matrix has a sum of 1. From the matrices of the individual participants, it is possible to obtain the average transition probability matrix of all participants belonging to the same subpopulation using the following equation, which must be applied for each training set available:

$$P(i \rightarrow j) = \frac{count_{ij}}{\sum_k count_{ik}}$$

[7]

Where $P(i \rightarrow j)$ is the transition probability from state $i$ to state $j$, $count_{ij}$ is the number of transitions from $i$ to $j$ and $\sum_k count_{ik}$ is the total number of transitions from state $i$.

3. Results and discussion

3.1. Inertia in the test responses for students with probable ADHD and probable OCD

The comparison of the degree of inertia in the transition matrices and the comparison of the stationary matrices in the two subgroups previously identified are described. Consider the two different subpopulations available as training sets. The procedure involves counting the 15 transitions that are distributed in the 5x5 square matrix, consecutively for each participant, and calculating the probabilities of each cell of the matrix in proportion to the set of transitions from each current state (in row) to itself and all other states. These probabilities constitute the basic parameters of the Markov chain model of the starting training set. In [8] are the transition probability matrices for the training sets of students with probable ADHD (n = 73, total number of transitions = 1095) and the group of students with probable OCD (n = 27; total number of transitions = 405).

Transition Probability Matrix (students with prob. OCD)

|  | | State j | | | | |
|---|---|---|---|---|---|---|
|  | | 1 | 2 | 3 | 4 | 5 |
| State i | 1 | 0.33 | 0.29 | 0.29 | 0.06 | 0.03 |
|  | 2 | 0.10 | 0.34 | 0.40 | 0.14 | 0.02 |
|  | 3 | 0.07 | 0.24 | 0.43 | 0.21 | 0.05 |
|  | 4 | 0.06 | 0.10 | 0.37 | 0.34 | 0.13 |
|  | 5 | 0.04 | 0.04 | 0.26 | 0.40 | 0.26 |

Transition Probability Matrix (students with prob. ADHD)

|  | | State j | | | | |
|---|---|---|---|---|---|---|
|  | | 1 | 2 | 3 | 4 | 5 |
| State i | 1 | 0.30 | 0.30 | 0.30 | 0.08 | 0.02 |
|  | 2 | 0.19 | 0.29 | 0.40 | 0.11 | 0.01 |
|  | 3 | 0.09 | 0.27 | 0.45 | 0.17 | 0.02 |
|  | 4 | 0.09 | 0.16 | 0.45 | 0.24 | 0.06 |
|  | 5 | 0.03 | 0.18 | 0.35 | 0.26 | 0.18 |

[8]

As stated before, the elements of the matrix main diagonal represent the probability of remaining in the same state in two consecutive items of the test. As already mentioned, this condition is called the inertia of the matrix. We expect a greater overall inertia in that subgroup whose disorder is more characterized by an active search for consistency in providing responses, which is assumed to be the group of students with probable OCD. To answer this question, and not being interested in studying the five cells of the main diagonal but only the set of transitions $s_{ii}$ to be compared to that of transitions $s_{ij}$, we will use the chi square for the association between the membership in the training set of students (with probable ADHD and probable OCD) and the relative number of inertial transitions (on diagonal) and of not inertial transitions (off the diagonal). In case of a significant test, the standardized residuals of the different cells are interpreted (criterion for std. residuals > 2). To proceed with this calculation, the transition probabilities $p_{ii}$ can be multiplied by the total number of transitions in the row.

$$\sum_{i=1}^{s} p_{ii} f_{i.}$$

[9]

Where s are the states, $p_{ii}$ are the transition probabilities that express the permanence in the same state in consecutive items, and $f_{i.}$ are the cumulative frequencies of the i-th rows through all the states in the subsequent event. From the inspection of the elements on the main diagonals in [8], it emerges that there is non-zero inertia in both groups. The most marked difference concerns the last two cells on the diagonal, even if a direct cell-by-cell comparison cannot be easily addressed because the sum of probabilities on the diagonal is not unitary. It is possible, however, to calculate the association statistic between the observed frequencies of inertial transitions and non-inertial transitions in the two intended subpopulations. The $\chi^2$ was not significant ($\chi^2(1) = 0.69$; p=0.41). Therefore, there does not seem to be a marked difference between the two training sets.

### 3.2. Stationary distributions in transition probability matrix for students with probable ADHD and probable OCD

The stationary probability matrix expresses the long-term behavior of the Markov chain. It represents the probability distribution that the Markov chain will take on over time if it is left to evolve independently and, in theory, infinitely. In other words, if a Markov chain has a stationary probability matrix, then the probability of being in a specific state will converge to the probability indicated by the stationary probability matrix, regardless of which state you start in. In this way, the stationary matrix expresses the probability of being in one of the predetermined states without any further constraints on previous states. The powers of the transition probability matrices can be calculated consecutively to obtain the stationary matrix after a certain number of steps. Different transition probability matrices converge with different powers. The stationary probability matrices of Markov chains do not converge all with the same power because convergence depends on the characteristics of the Markov chain transition matrix. In general, it can be shown that convergence with power elevation exists if and only if the transition matrix is irreducible and aperiodic. A transition matrix is irreducible if, for each pair of states *(i,j)*, there exists a finite path from *i* to *j*. A transition matrix is aperiodic if the least common multiple of the periods of all states is 1. The transition matrices of our study are irreducible and aperiodic (see Table 1).

Table 1. first column reports the power applied to the corresponding transition probability matrix. The other columns are devoted to report matrices for the two groups of students across successive powers.

|    | Powers of the transition probability matrix (students with prob. ADHD) | | | | | Powers of the transition probability matrix (students with prob. OCD) | | | | |
|----|-------|-------|-------|-------|-------|-------|-------|-------|-------|-------|
|    | 0.300 | 0.300 | 0.300 | 0.080 | 0.020 | 0.330 | 0.290 | 0.290 | 0.060 | 0.030 |
|    | 0.190 | 0.290 | 0.400 | 0.110 | 0.010 | 0.100 | 0.340 | 0.400 | 0.140 | 0.020 |
| P¹ | 0.090 | 0.270 | 0.450 | 0.170 | 0.020 | 0.070 | 0.240 | 0.430 | 0.210 | 0.050 |
|    | 0.090 | 0.160 | 0.450 | 0.240 | 0.060 | 0.060 | 0.100 | 0.370 | 0.340 | 0.130 |
|    | 0.030 | 0.180 | 0.350 | 0.260 | 0.180 | 0.040 | 0.040 | 0.260 | 0.400 | 0.260 |
|    | 0.182 | 0.274 | 0.388 | 0.132 | 0.023 | 0.163 | 0.271 | 0.366 | 0.154 | 0.046 |
|    | 0.158 | 0.269 | 0.406 | 0.144 | 0.023 | 0.104 | 0.255 | 0.394 | 0.193 | 0.053 |
| P² | 0.135 | 0.258 | 0.421 | 0.159 | 0.027 | 0.092 | 0.228 | 0.392 | 0.220 | 0.069 |
|    | 0.121 | 0.244 | 0.423 | 0.175 | 0.038 | 0.081 | 0.179 | 0.376 | 0.263 | 0.100 |
|    | 0.104 | 0.230 | 0.419 | 0.191 | 0.057 | 0.070 | 0.138 | 0.355 | 0.303 | 0.135 |
|    | 0.154 | 0.264 | 0.407 | 0.149 | 0.026 | 0.118 | 0.245 | 0.382 | 0.195 | 0.061 |
| P³ | 0.149 | 0.262 | 0.411 | 0.152 | 0.027 | 0.101 | 0.233 | 0.387 | 0.212 | 0.067 |
|    | 0.142 | 0.259 | 0.414 | 0.156 | 0.028 | 0.096 | 0.223 | 0.385 | 0.222 | 0.073 |
|      | 0.138 | 0.256 | 0.416 | 0.160 | 0.031 | 0.091 | 0.205 | 0.380 | 0.238 | 0.085 |
|      | 0.131 | 0.252 | 0.417 | 0.165 | 0.035 | 0.085 | 0.188 | 0.375 | 0.255 | 0.097 |
|      | 0.147 | 0.261 | 0.411 | 0.153 | 0.028 | 0.104 | 0.231 | 0.384 | 0.212 | 0.069 |
|      | 0.146 | 0.261 | 0.412 | 0.154 | 0.028 | 0.099 | 0.225 | 0.385 | 0.219 | 0.072 |
| $P^4$ | 0.144 | 0.260 | 0.413 | 0.155 | 0.028 | 0.097 | 0.221 | 0.384 | 0.223 | 0.075 |
|      | 0.143 | 0.259 | 0.414 | 0.156 | 0.029 | 0.095 | 0.215 | 0.382 | 0.229 | 0.079 |
|      | 0.141 | 0.258 | 0.414 | 0.158 | 0.030 | 0.092 | 0.208 | 0.381 | 0.236 | 0.083 |
|      | 0.145 | 0.260 | 0.412 | 0.154 | 0.028 | 0.100 | 0.225 | 0.384 | 0.219 | 0.072 |
|      | 0.145 | 0.260 | 0.412 | 0.155 | 0.028 | 0.098 | 0.222 | 0.384 | 0.221 | 0.074 |
| $P^5$ | 0.145 | 0.260 | 0.413 | 0.155 | 0.028 | 0.097 | 0.221 | 0.384 | 0.223 | 0.075 |
|      | 0.144 | 0.260 | 0.413 | 0.155 | 0.028 | 0.096 | 0.218 | 0.383 | 0.226 | 0.077 |
|      | 0.144 | 0.259 | 0.413 | 0.156 | 0.028 | 0.095 | 0.216 | 0.383 | 0.228 | 0.078 |
|      | 0.145 | 0.260 | 0.412 | 0.155 | 0.028 | 0.098 | 0.222 | 0.384 | 0.221 | 0.074 |
|      | 0.145 | 0.260 | 0.412 | 0.155 | 0.028 | 0.098 | 0.221 | 0.384 | 0.223 | 0.075 |
| $P^6$ | 0.145 | 0.260 | 0.413 | 0.155 | 0.028 | 0.097 | 0.221 | 0.384 | 0.223 | 0.075 |
|      | 0.145 | 0.260 | 0.413 | 0.155 | 0.028 | 0.097 | 0.220 | 0.383 | 0.224 | 0.076 |
|      | 0.144 | 0.260 | 0.413 | 0.155 | 0.028 | 0.097 | 0.219 | 0.383 | 0.225 | 0.076 |
|      | 0.145 | 0.260 | 0.412 | 0.155 | 0.028 | 0.098 | 0.221 | 0.384 | 0.223 | 0.075 |
|      | 0.145 | 0.260 | 0.412 | 0.155 | 0.028 | 0.098 | 0.221 | 0.384 | 0.223 | 0.075 |
| $P^7$ | 0.145 | 0.260 | 0.412 | 0.155 | 0.028 | 0.097 | 0.221 | 0.384 | 0.223 | 0.075 |
|      | 0.145 | 0.260 | 0.413 | 0.155 | 0.028 | 0.097 | 0.220 | 0.384 | 0.224 | 0.075 |
|      | 0.145 | 0.260 | 0.413 | 0.155 | 0.028 | 0.097 | 0.220 | 0.383 | 0.224 | 0.076 |
|      |       |       |       |       |       | 0.098 | 0.221 | 0.384 | 0.223 | 0.075 |
|      |       |       |       |       |       | 0.098 | 0.221 | 0.384 | 0.223 | 0.075 |
| $P^8$ |       |       |       |       |       | 0.097 | 0.221 | 0.384 | 0.223 | 0.075 |
|      |       |       |       |       |       | 0.097 | 0.221 | 0.384 | 0.223 | 0.075 |
|      |       |       |       |       |       | 0.097 | 0.220 | 0.384 | 0.223 | 0.075 |
|      |       |       |       |       |       | 0.098 | 0.221 | 0.384 | 0.223 | 0.075 |
|      |       |       |       |       |       | 0.098 | 0.221 | 0.384 | 0.223 | 0.075 |
| $P^9$ |       |       |       |       |       | 0.097 | 0.221 | 0.384 | 0.223 | 0.075 |
|      |       |       |       |       |       | 0.097 | 0.221 | 0.384 | 0.223 | 0.075 |
|      |       |       |       |       |       | 0.097 | 0.221 | 0.384 | 0.223 | 0.075 |
|      |       |       |       |       |       | 0.098 | 0.221 | 0.384 | 0.223 | 0.075 |
|      |       |       |       |       |       | 0.097 | 0.221 | 0.384 | 0.223 | 0.075 |
| $P^{10}$ |    |       |       |       |       | 0.097 | 0.221 | 0.384 | 0.223 | 0.075 |
|      |       |       |       |       |       | 0.097 | 0.221 | 0.384 | 0.223 | 0.075 |
|      |       |       |       |       |       | 0.097 | 0.221 | 0.384 | 0.223 | 0.075 |
|      |       |       |       |       |       | 0.097 | 0.221 | 0.384 | 0.223 | 0.075 |
|      |       |       |       |       |       | 0.097 | 0.221 | 0.384 | 0.223 | 0.075 |
| $P^{11}$ |    |       |       |       |       | 0.097 | 0.221 | 0.384 | 0.223 | 0.075 |
|      |       |       |       |       |       | 0.097 | 0.221 | 0.384 | 0.223 | 0.075 |
|      |       |       |       |       |       | 0.097 | 0.221 | 0.384 | 0.223 | 0.075 |

The sum of the probabilities for the different states is equal to 1. When two or more matrices from different training sets of data reach convergence, then it is possible to compare their structure, in this case adopting a chi-squared goodness-of-fit analysis model - as suggested elsewhere by Visser, Raijmakers and Molenaar (2002) for assessing the fit of models of Hidden Markov Chains applied to psychological data. In case of a significant test, the standardized residuals of the different cells are interpreted (criterion for std. residuals > 2). The two matrices of transitions for students with probable ADHD and probable OCD converge to the seventh and eleventh power, respectively. In this case, it is easy to compare the two vectors (see Table 2).

The goodness-of-fit between the stationary matrix of the training set of participants with probable ADHD (reference group, conventionally the expected frequencies) and that of the training set of participants with probable OCD (focal group, conventionally the observed frequencies) was found to be significant: $\chi^2(4)=57.35$; $p<0.001$. Therefore, the data set of the focal group does not fit that of the reference group. From the inspection of the standardized residuals, the probability of transition relative to state 4 and state 5 is greater than the criterion. Therefore, in the long term, there is a greater probability of transition on states associated with high values by the group of students with probable OCD than those with probable ADHD.

### 3.3. Fitting the sequences of participants with different Markov chain models

As mentioned above, the second main goal of the study is to be able to achieve two Markov chain models and resolve which one is the most compatible with a certain sequence. In other words, values for a likelihood ratio test are needed. In the present study two Markov chain models are generated corresponding to the two populations of students with probable OCD and Probable ADHD (Og and Ag groups, respectively). The transition probabilities for each one of these two models are obtained through these equations:

$$P_{ij}^{A_g} = \frac{count_{ij}^{A_g}}{\sum_k count_{ik}^{A_g}}$$

[10]

$$P_{ij}^{O_g} = \frac{count_{ij}^{O_g}}{\sum_k count_{ik}^{O_g}}$$

[11]

They can be considered the maximum likelihood (ML) estimators for the transition probabilities of each training set. These two models can be used to calculate a score that can discriminate whether a given sequence is most likely compatible with the transition matrix of one or the other of the two training sets. The two models can be integrated into the following equation, as in Paxinou et al. (2021):

$$Score(x) = \log_2 \frac{(x|model^{O_g})}{(x|model^{A_g})} = \sum_{k=1}^{N} \frac{p_{x_{k-1}x_k}^{O_g}}{p_{x_{k-1}x_k}^{A_g}} = \sum_{k=1}^{N} \beta_{x_{k-1}x_k}$$

[12]

where $\beta_{x_{k-1}x_k}$ are the log likelihood ratios for each corresponding transition probability within the matrices. The transformation to logarithm improves the practicality, numerical stability, and interpretation of the estimates obtained through the maximization of the likelihood in the score. It is important to note that a positive score expresses a greater compatibility of the participant's sequence with the transition probability model at the numerator. On the other hand, a negative score expresses a greater compatibility between the sequence under evaluation and the model in the denominator. Zero, therefore, represents the cut-off score for evaluating the membership in one of the two groups of the sequence under study. The first step, therefore, consists in the construction of a matrix of the ratio, element by element, between the two available transition probability matrices: $R = P_{O_g}/P_{A_g}$.

|  |  | State j |  |  |  |  |
|---|---|---|---|---|---|---|
|  |  | 1 | 2 | 3 | 4 | 5 |
| State i | 1 | 1.10 | 0.97 | 0.97 | 0.75 | 1.50 |
|  | 2 | 0.53 | 1.17 | 1.00 | 1.27 | 2.00 |
|  | 3 | 0.78 | 0.89 | 0.96 | 1.24 | 2.50 |
|  | 4 | 0.67 | 0.63 | 0.82 | 1.42 | 2.17 |
|  | 5 | 1.33 | 0.22 | 0.74 | 1.54 | 1.44 |

[13]

The second step consists in the transformation, element by element, into the log2 of the obtained matrix: $LR = Log_2\left(P_{O_g}/P_{A_g}\right)$. By summing the product of the frequency with which each transition appears in the sequence of a student and the relative log-ratio value, we obtain the score for that student.

|  |  | State j |  |  |  |  |
|---|---|---|---|---|---|---|
|  |  | 1 | 2 | 3 | 4 | 5 |
| State i | 1 | 0.14 | -0.04 | -0.04 | -0.42 | 0.58 |
|  | 2 | -0.92 | 0.23 | 0.00 | 0.34 | 1.00 |
|  | 3 | -0.36 | -0.17 | -0.06 | 0.31 | 1.32 |
|  | 4 | -0.58 | -0.67 | -0.29 | 0.51 | 1.12 |
|  | 5 | 0.41 | -2.18 | -0.43 | 0.62 | 0.53 |

[14]

This score can be used as a classification in the way already described previously: positive scores express compatibility with the model in the numerator of the ratio, negative scores denote a compatibility of the sequence with the model in the denominator. Considering again the sequence of student O05 (3243232443244333) it is easy to observe that the sequence is composed of: one transition "2-3", three transitions "2-4", four transitions "3-2", two transitions "3-3", three transitions "4-3" and two transitions "4-4". Then:

$$Score(O05) = (1 \cdot 0.00) + (3 \cdot 0.34) + (4 \cdot -0.17) + (2 \cdot -0.06) + (3 \cdot -0.29) + (2 \cdot 0.51) = 0.37$$

[15]

The positive score for student O05 indicates that his sequence has a greater compatibility with the model at the numerator, which refers to the training set of students with probable OCD. In fact, he/she had been classified through the independent assessment (see the method section) in the group of probable OCD. Assuming that all students are reclassified (cut-off = 0) through the score described, it is possible to show that the reclassification obtained has adequate operating characteristics and positive and negative Likelihood ratio (LR+, LR-, respectively) useful on a diagnostic level, as a possible supplement to what is already available in the assessment. In the following table 2 and figure 1, the diagnostic value of the score based on the Markov chain and of the sum score of the test compared each other.

Table 2. Diagnostic accuracy of Markov Chain (Score) calculated as log-likelihood ratio between the model of probable OCD and probable ADHD students. TP: true positives, FN: false negatives (conventionally probable OCD students), TN: true negatives and FP: false positives (conventionally probable ADHD students), Sn is the sensitivity, Sp is the specificity, LR+ is the positive likelihood ratio (Sn/(1-Sp)), LR- is the negative likelihood ratio ((1-Sn)/ Sp). AUC is the area under the curve.

| Cut-off | TP | FN | TN | FP | Sn | Sp | LR+ | LR- | AUC |
|---|---|---|---|---|---|---|---|---|---|
| 4FMW total score (>= 44) | 17 | 10 | 38 | 35 | 0.63 | 0.52 | 1.31 | 0.71 | 0.612 |
| Likelihood log ratio (>=0) | 21 | 6 | 56 | 17 | 0.78 | 0.77 | 3.34 | 0.29 | 0.753 |

Interpreting the LR+, we can say that a score - obtained using the described procedure - greater than 0 is 3.34 times more likely in students with probable OCD than in those with probable ADHD. Considering for comparison the best cut-off of the total scale of the 4FMW, a score greater than or equal to 44 is only 1.31 times more likely in students with probable OCD than in those with probable ADHD. This comparison can also be appreciated graphically in the (1-specificity)/sensitivity graph, which allows to visualize the area under the curve in the two scores employed.

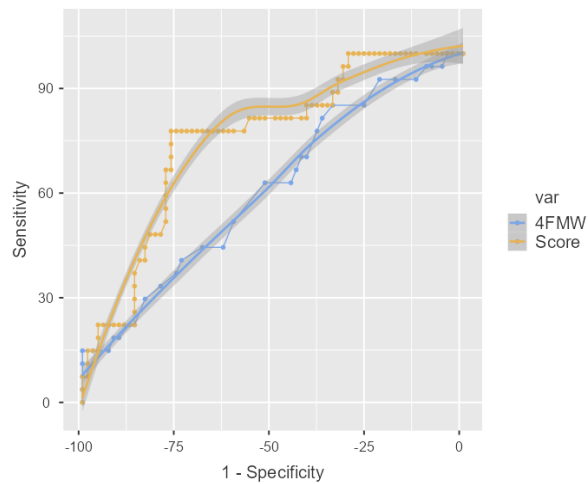

Figure 1. Comparison of the ROC curves based on the Markov chain score and the sum score of the test

4. Compare the dynamic of test responses with Markov Chain Models based on theorical or simulated distributions of transition probabilities

It may happen that there are no data on subgroups to be differentiated, but that it is still desired to evaluate whether there is any pattern in the dynamic of test response sequences of a sample of participants. We can examine the response dynamics of a sample of participants to search for patterns in their test-taking behavior, even in the absence of subgroup data for comparison. Markov chains can also be used to compare observed sequences with theoretical or simulated models that can still provide scores to be used as a classification criterion. Then, it is possible to construct theoretical matrices of transition probabilities or samples of known size that follow predefined discrete probability distributions with which to inform the Markov Chain model.

First, a comparison between the transitions reflecting a stochastic process characterized by a stated structure in the data (i.e., the drunkard's walk model, DWM) and one in which the transitions from any state to any other state is equally probable (max entropy model, MEM) is developed. The drunkard's walk scenario is characterized by a series of consecutive steps in which one proceeds from a starting position to an ending position by moving right-forward or left-forward but not more than one position at a time. In some cases, a step forward without moving to the right or left is performed, remaining in line. In the context of drunkard's walk scenario, the transition matrix represents the probabilities of moving from one position to another in discrete time steps. As we know, each row of the matrix corresponds to the current state (or position), and each column corresponds to the possible afterward states. The values in the matrix identify the probabilities of moving from one state to another. In the simple one-dimensional drunkard's walk, the diagonal elements of the transition matrix represent the probability of remaining in line, and the off-diagonal elements represent the probabilities of moving left or right, no more than one step at time. In other word, any transitions regarding more than one step, e.g., 1-3, 2-5 etc. should be represented in the matrix, in principle, with a negligible probability. An example of a transition probability matrix for a drunkard's walk-like scenario with five states / positions compared with a matrix showing all equally probable transitions follow in [16] and [17], respectively:

|   |   | State j |   |   |   |   |
|---|---|---|---|---|---|---|
|   |   | 1 | 2 | 3 | 4 | 5 |
| State i | 1 | 0.50 | 0.47 | 0.01 | 0.01 | 0.01 |
|   | 2 | 0.24 | 0.50 | 0.24 | 0.01 | 0.01 |
|   | 3 | 0.01 | 0.24 | 0.50 | 0.24 | 0.01 |
|   | 4 | 0.01 | 0.01 | 0.24 | 0.50 | 0.24 |
|   | 5 | 0.01 | 0.01 | 0.01 | 0.47 | 0.50 |

[16]

|   |   | State j |   |   |   |   |
|---|---|---|---|---|---|---|
|   |   | 1 | 2 | 3 | 4 | 5 |
| State i | 1 | 0.20 | 0.20 | 0.20 | 0.20 | 0.20 |
|   | 2 | 0.20 | 0.20 | 0.20 | 0.20 | 0.20 |
|   | 3 | 0.20 | 0.20 | 0.20 | 0.20 | 0.20 |
|   | 4 | 0.20 | 0.20 | 0.20 | 0.20 | 0.20 |
|   | 5 | 0.20 | 0.20 | 0.20 | 0.20 | 0.20 |

[17]

In the [16] the probability of staying in the same position is 0.5 (the diagonal elements of the matrix). The probability of moving one step to the right is 0.47 for the upper left off-diagonal element (that is, moving from 1 to 2). The probability of moving one step to the left is 0.47 for the lower right off-diagonal element (that is, moving from 5 to 4). The probability of moving one step to the left or to the right of the state i from 2 to 4 is 0.24. The probability for all the other off-diagonal elements is equal to 1/100. This small probability value is necessary for calculation reasons, since a ratio value of zero would make the transformation into $\log_2$ unfeasible.

This matrix enables researchers to investigate an example of the so-called drunkard's walk model, specifically focusing, for instance, on whether it converges to steady state distribution. More interestingly for the psychological testing theory, it allows to know how many sequences showed this peculiar pattern in test responses. It could, for example, reveal intentionally distorting behaviors as well as highlight entirely unaware tendencies - such as hysteresis

- briefly described in the introduction section - linked to the influence that responding to the test itself has on the participants and which could be reflected in state choice, one item after another. Examples of the drunkard's walk models have been used in psychology and neuroscience to describe eye movements (Rucci & Victor, 2015), neural firing cascade models (Martinello et al., 2017), decision-making times and outcomes (Nosofsky & Palmeri, 1997), predictions of human motion flows in the environment (Jiang, 2009).

Nonetheless, other models can be used to define different matrices. For example:

• people can favor central states with the most likely responses occurring in the center of the scale in a symmetrical way (symmetric),

• People can favor the first part of the scale. The distribution of frequences is characterized by a tail on the right side of the graph, with the most likely responses occurring at the lower end of the scale - typical of cases in which a test that characterizes a pathological state is administered to a sample randomly drawn from the general population (skewed+),

• People can favor the last part of the scale. The distribution of frequences is characterized by a tail on the left side of the graph, indicating that most responses cluster at the higher end of the scale. This type of probabilities can arise when a test designed to assess a pathological state is given to a clinical population (skewed-). An example of the discrete stationary transition vectors associated with each model follow (table 3).

Table 3. Probabilities for three stationary transition matrices according to 1) a symmetric, 2) a positive skewed, and 3) a negative skewed models of state probabilities

|           | $p_{i1}$ | $p_{i2}$ | $p_{i3}$ | $p_{i4}$ | $p_{i5}$ |
|-----------|------|------|------|------|------|
| Symmetric | 0.10 | 0.20 | 0.40 | 0.20 | 0.10 |
| Skewed+   | 0.25 | 0.40 | 0.20 | 0.10 | 0.05 |
| Skewed-   | 0.05 | 0.10 | 0.20 | 0.40 | 0.25 |

Each stationary transition vector describes the probability of participants giving a particular response for each state. Each of them is compared in a log-likelihood ratio with the matrix based on maximum entropy (MEM). If none of the previous distributions can adequately describe the dynamics of the test responses of a given participant, then it is likely that the matrix uniformly distributing the transition probabilities may have the highest likelihood for the participant's data, which would contribute to the hypothesis that the responses were provided not considering previous responses. On the contrary, the other three distributions would express three different dynamics in giving the responses, all plausible, in different contexts of the assessment practice.

In the case in which sequences of participants drawn from the normal population and without asserted psychological disorders are evaluated in a test that intercepts symptoms, i.e., of mind wandering, the sequences should be compatible mainly with the "skewed+" and in second instance with the "symmetric" models, respectively. Conversely, they should not be compatible with the matrix of transition probabilities generated by the "skewed-" model and less likely with that generated by the maximum entropy model reported in [17].

4.1. Materials and Methods

The assessment procedure described in section 2 and 2.1. makes available a subsample of students without any sign of mental disorders. One-hundred-and-eighty students (142 women; mean age = 21.66, SD = 1.61) have taken part to the study with the same procedure as described before. The materials are the same as in the previous study.

## 5. Results and Discussion

### 5.1. Theorical distributions of transition probabilities: The drunkard's walk scenario

The likelihood ratio and log2 of ratio matrices are reported in [18] and [19], respectively.

|  |  | State j |  |  |  |  |
|--|--|---|---|---|---|---|
|  |  | *1* | *2* | *3* | *4* | *5* |
| State i | *1* | 2.5 | 2.35 | 0.05 | 0.05 | 0.05 |
|  | *2* | 1.20 | 2.50 | 1.20 | 0.05 | 0.05 |
|  | *3* | 0.05 | 1.20 | 2.50 | 1.20 | 0.05 |
|  | *4* | 0.05 | 0.05 | 1.20 | 2.50 | 1.20 |
|  | *5* | 0.05 | 0.05 | 0.05 | 2.35 | 2.50 |

[18]

|  |  | State j |  |  |  |  |
|--|--|---|---|---|---|---|
|  |  | *1* | *2* | *3* | *4* | *5* |
| State i | *1* | 1.32 | 1.23 | -4.32 | -4.32 | -4.32 |
|  | *2* | 0.26 | 1.32 | 0.26 | -4.32 | -4.32 |
|  | *3* | -4.32 | 0.26 | 1.32 | 0.26 | -4.32 |
|  | *4* | -4.32 | -4.32 | 0.26 | 1.32 | 0.26 |
|  | *5* | -4.32 | -4.32 | -4.32 | 1.23 | 1.32 |

[19]

Following the calculation of students' scores based on the ratio between DWM and MEM models, it emerges that a half of the sample obtained a positive result. Most of them are characterized by a significant level of inertia. The drunkard's walk transition probabilities seem associated to a dynamic of responses in which the immediately previous choice appear to anchor the subsequent one. This determines an increase in inertial responses, those in which one remains in the same response category, or in the immediately neighboring categories.

### 5.2. Compare the dynamic of test responses with three theorical stationary matrices

The likelihood ratios between the transition matrix for the symmetric, the positively and negatively skewed arrangements of transition probabilities with the maximum entropy model [17] as a reference are calculated, as follow in [20], [21] and [22], respectively.

|  | | State j | | | | |
|---|---|---|---|---|---|---|
|  | | 1 | 2 | 3 | 4 | 5 |
| State i | 1 | 0.50 | 1.00 | 2.00 | 1.00 | 0.50 |
|  | 2 | 0.50 | 1.00 | 2.00 | 1.00 | 0.50 |
|  | 3 | 0.50 | 1.00 | 2.00 | 1.00 | 0.50 |
|  | 4 | 0.50 | 1.00 | 2.00 | 1.00 | 0.50 |
|  | 5 | 0.50 | 1.00 | 2.00 | 1.00 | 0.50 |

[20]

|  | | State j | | | | |
|---|---|---|---|---|---|---|
|  | | 1 | 2 | 3 | 4 | 5 |
| State i | 1 | 1.25 | 2.00 | 1.00 | 0.50 | 0.25 |
|  | 2 | 1.25 | 2.00 | 1.00 | 0.50 | 0.25 |
|  | 3 | 1.25 | 2.00 | 1.00 | 0.50 | 0.25 |
|  | 4 | 1.25 | 2.00 | 1.00 | 0.50 | 0.25 |
|  | 5 | 1.25 | 2.00 | 1.00 | 0.50 | 0.25 |

[21]

|  | | State j | | | | |
|---|---|---|---|---|---|---|
|  | | 1 | 2 | 3 | 4 | 5 |
| State i | 1 | 0.25 | 0.50 | 1.00 | 2.00 | 1.25 |
|  | 2 | 0.25 | 0.50 | 1.00 | 2.00 | 1.25 |
|  | 3 | 0.25 | 0.50 | 1.00 | 2.00 | 1.25 |
|  | 4 | 0.25 | 0.50 | 1.00 | 2.00 | 1.25 |
|  | 5 | 0.25 | 0.50 | 1.00 | 2.00 | 1.25 |

[22]

Then, the previous three matrices have been transformed, element by element, into their respective log2s, as usual They are in [23], [24] and [25], respectively.

|  | | State j | | | | |
|---|---|---|---|---|---|---|
|  | | 1 | 2 | 3 | 4 | 5 |
| State i | 1 | -1.00 | 0.00 | 1.00 | 0.00 | -1.00 |
|  | 2 | -1.00 | 0.00 | 1.00 | 0.00 | -1.00 |
|  | 3 | -1.00 | 0.00 | 1.00 | 0.00 | -1.00 |
|  | 4 | -1.00 | 0.00 | 1.00 | 0.00 | -1.00 |
|  | 5 | -1.00 | 0.00 | 1.00 | 0.00 | -1.00 |

[23]

|   | | State j | | | | |
|---|---|---|---|---|---|---|
|   |   | 1 | 2 | 3 | 4 | 5 |
| State i | 1 | 0.32 | 1.00 | 0.00 | -1.00 | -2.00 |
|   | 2 | 0.32 | 1.00 | 0.00 | -1.00 | -2.00 |
|   | 3 | 0.32 | 1.00 | 0.00 | -1.00 | -2.00 |
|   | 4 | 0.32 | 1.00 | 0.00 | -1.00 | -2.00 |
|   | 5 | 0.32 | 1.00 | 0.00 | -1.00 | -2.00 |

[24]

|   | | State j | | | | |
|---|---|---|---|---|---|---|
|   |   | 1 | 2 | 3 | 4 | 5 |
| State i | 1 | -2.00 | -1.00 | 0.00 | 1.00 | 0.32 |
|   | 2 | -2.00 | -1.00 | 0.00 | 1.00 | 0.32 |
|   | 3 | -2.00 | -1.00 | 0.00 | 1.00 | 0.32 |
|   | 4 | -2.00 | -1.00 | 0.00 | 1.00 | 0.32 |
|   | 5 | -2.00 | -1.00 | 0.00 | 1.00 | 0.32 |

[25]

Three scores were calculated for each student, one for each comparison. Then the 180 students were classified according to the maximum likelihood of their sequence with one of the models adopting the following algorithm:

If

score[symm/MEM] AND score [skewed+/MEM] AND score [skewed-/MEM] < 0

Then

MEM

Else

the highest positive score among the three scores calculated suggests the most probable membership.

A chi-square test was applied with the null hypothesis of equiprobable frequencies. It is significant ($\chi^2(3)=140.58$; $p<0.001$). As expected, the participants with positively skewed (a predominance of states 1 and 2) and symmetric (a predominance of central states) transition probabilities are prevalent. All participants with the highest values in the total test scale (range 57-72) also have a higher fit with the negatively skewed distribution, on the contrary, all those with the lowest values (16-35) show the greatest fit with the positively skewed model. Interestingly, students with total scores immediately lower than that belonging to the highest range observed, that is, 55 and 56 (n=7), two of them showed the maximum likelihood with the MEM suggesting that they responded without a defined pattern (among those tested). Three students whose sequence is more fitting with the symmetric model than with the expected "skewed-". In the central part of the distribution (36-54), on the other hand, at the same test score, we find both students fitting the "symmetric" as well as "skewed+" models, respectively and this suggests that within the sample of respondents without recognized signs of mental disease, there are different response dynamics, that make possible to infer different response styles, for instance, oriented to remain in low or central states.

6. Conclusions

The results of the present study suggest that Markov chain models can be used to effectively characterize the dynamics of test responses. By comparing the transition matrices of different groups of respondents, in terms of inertia, stationary distribution and scores based on the log-likelihood obtained by the ratio, element by element, between matrices, it is possible to identify differences in their response styles. These differences can be related to a variety of factors, including individual differences, impression management and the features of the test, themselves. In the present study, a log-likelihood ratio test was used to compare the transition matrices of two groups of respondents, students with probable OCD or probable ADHD. The results showed that the two groups had significantly different stationary distribution. Moreover, the score seems to be able to discriminate the two groups based on their transition matrices better than does the sum scale of the test. This could help to identify the characteristic response styles of people with different psychological disorders and integrate the differential diagnosis with a new source of information.

In addition to comparing the transition matrices of different groups of respondents, Markov chain models can also be used to compare the response sequences of a single group of respondents to theoretical transition matrices. This could help to identify the extent to which the respondents' answers are consistent with a particular theoretical model.

The concepts of autocorrelation, hysteresis, and path dependency are all relevant to the study of Markov chain models. Autocorrelation refers to the correlation between the responses of a single respondent at different time points. Hysteresis refers to the phenomenon in which the current state of a system is influenced by its past states. Path dependency refers to the phenomenon in which the current state of a system is influenced by its initial state. In the context of psychological testing, autocorrelation could be used to measure the consistency of participants' responses over time. Hysteresis could be referred to measure the extent to which a respondent's responses are affected by their previous responses. Path dependency could be referred to measure the extent to which a respondents' responses are affected by their initial state.

Markov matrices can be employed to model and analyze responses to multiple-choice psychological scales in various ways. Here are some approaches that one could consider: a) modeling transitions between responses. A transition matrix can be constructed to describe the probabilities of transitioning between responses. For instance, if there is a 5-point scale ranging from "strongly agree" to "strongly disagree," the Markov matrix could indicate the probabilities of transitioning from one response to another, b) analysis of response patterns. Once the Markov matrix is constructed, one can analyze individuals' response patterns. For example, common sequences of responses or transition patterns that occur frequently could be identified. This may help in recognizing trends in how people respond to questions or in identifying subgroups with distinct response patterns, c) Prediction of future responses. Given a sequence of past responses, the probability of future responses can be calculated, d) assessment of the internal consistency. The Markov matrix can also be utilized to assess the internal consistency of individuals' responses. For instance, one could calculate the autocorrelation of responses using the Markov model to determine whether successive responses are correlated with each other, which may indicate internal consistency or a lack thereof in how a person responds to the test, and e) evaluation of test effectiveness. The Markov matrix can also be employed to evaluate the effectiveness of the psychological test in measuring the construct it aims to assess. For example, observed transitions in individuals' responses could be compared with transitions predicted by the Markov model to assess whether the test accurately measures what it intends to measure.

In summary, Markov matrices can be used to thoroughly analyze responses to multiple-choice psychological scales, enabling the identification of patterns, making predictions, and evaluating test effectiveness.

These are just a few examples of how Markov chain models can be used to study testing and survey responses. As research in this area continues, it is likely that we will learn more about the ways in which these models can be used to improve our understanding of the dynamics of test responses. The present study has suggested the potential of Markov chain models for the study of testing.

In conclusion by using Markov chain models, it is possible to gain a deeper understanding of the dynamics of test responses and the factors that influence them. Moreover, the application of the model can help in identifying common response patterns, predicting an individual's future responses based on their previous responses in a test. Future applications of the model in psychological testing can be useful to detect potential inconsistencies as well as random responses.

Acknowledgments

This research received no external funding.

The study was based on retrospective data collected for the following primary studies: Lopez, A., Caffò, A. O., Tinella, L., Di Masi, M. N., & Bosco, A. (2021). Variations in mindfulness associated with the COVID-19 outbreak: Differential effects on cognitive failures, intrusive thoughts and rumination. *Applied Psychology: Health and Well-Being*, 13(4), 761-780, and Lopez, A., Caffò, A. O., Tinella, L., & Bosco, A. (2023). The four factors of mind wandering questionnaire: Content, construct, and clinical validity. *Assessment*, 30(2), 433-447. The Ethical Committee of the Institution approved the general protocol of both studies (n. 3660-CEL03/17). They were performed following the Helsinki Declaration and its later amendments.

Informed consent was obtained from all participants involved in the two primary studies.

The Author would like to thank Antonella Lopez, Giuseppina Spano and Luigi Tinella who read previous versions of the manuscript and for their fruitful suggestions.

The author declares no conflict of interest.

References

Ackerman, P. L., Kanfer, R., and Beier, M. E. (2013). Trait complex, cognitive ability, and domain knowledge predictors of baccalaureate success, STEM persistence, and gender differences. *Journal of Educational Psychology* 105:911. doi: 10.1037/a0032338

Atmanspacher, H., & Römer, H. (2012). Order effects in sequential measurements of non-commuting psychological observables. *Journal of Mathematical Psychology*, 56(4), 274-280.

Bindler, A., & Hjalmarsson, R. (2019). Path dependency in jury decision making. *Journal of the European Economic Association*, 17(6), 1971-2017.

Brainerd, C. J. (1979). Markovian interpretations of conservation learning. *Psychological Review*, 86(3), 181.

Cole, S.N., Tubbs, P.M.C. (2022). Predictors of obsessive–compulsive symptomology: mind wandering about the past and future. *Psychological Research,* 86, 1518–1534.

Doss, R. A., & Lowmaster, S. E. (2022). Validation of the DSM-5 Level 1 Cross-Cutting Symptom Measure in a Community Sample. *Psychiatry Research*, 318, 114935.

Du, Y., & Clark, J. E. (2017). New insights into statistical learning and chunk learning in implicit sequence acquisition. *Psychonomic bulletin & review*, 24, 1225-1233.

Fossati, A., Borroni, S., & Del Corno, F. (2015). *DSM-5: Scale di Valutazione Adulti*. Raffaello Cortina.

Hogarth, R. M., & Einhorn, H. J. (1992). Order effects in belief updating: The belief-adjustment model. *Cognitive psychology*, 24(1), 1-55.

Jiang, B. (2009). Ranking spaces for predicting human movement in an urban environment. *International Journal of Geographical Information Science*, 23(7), 823-837.

Karush, J. (1961). On the Chapman-kolmogorov equation. *The Annals of Mathematical Statistics*, 32(4), 1333-1337.

Kintsch, W., & Morris, C. J. (1965). Application of a Markov model to free recall and recognition. *Journal of Experimental Psychology*, 69(2), 200.

Lopez, A., Caffò, A. O., Tinella, L., & Bosco, A. (2023). The four factors of mind wandering questionnaire: Content, construct, and clinical validity. *Assessment*, 30(2), 433-447.

Madras, N., & Randall, D. (2002). Markov chain decomposition for convergence rate analysis. *Annals of Applied Probability*, 581-606.

Martinello, M., Hidalgo, J., Maritan, A., Di Santo, S., Plenz, D., & Muñoz, M. A. (2017). Neutral theory and scale-free neural dynamics. *Physical Review X*, 7(4), 041071.

McMillan, R. L., Kaufman, S. B., & Singer, J. L. (2013). Ode to positive constructive daydreaming. *Frontiers in Psychology*, 4, 626.

Migliarese, G., Venturi, V., Cerveri, G., & Mencacci, C. L. (2015). ADHD nell'adulto: misdiagnosi e incidenza della patologia nei servizi. *Psichiatria oggi*, 28, 16-25.

Nesterov, Y., & Nemirovski, A. (2015). Finding the stationary states of Markov chains by iterative methods. *Applied Mathematics and Computation*, 255, 58-65.

Nosofsky, R. M., & Palmeri, T. J. (1997). An exemplar-based random walk model of speeded classification. *Psychological review*, 104(2), 266.

Odic, D., Hock, H., & Halberda, J. (2014). Hysteresis affects approximate number discrimination in young children. *Journal of Experimental Psychology: General*, 143(1), 255.

Osborne, J. B., Zhang, H., Carlson, M., Shah, P., & Jonides, J. (2023). The association between different sources of distraction and symptoms of attention deficit hyperactivity disorder. *Frontiers in Psychiatry*, 14.

Paulhus, D. L. (1991). Measurement and control of response bias. In J. P. Robinson, P. R. Shaver, & L. S. Wrightsman (Eds.), *Measures of personality and social psychological attitudes* (pp. 17–59). Academic Press.

Paxinou, E., Kalles, D., Panagiotakopoulos, C. T., & Verykios, V. S. (2021). Analyzing sequence data with Markov chain models in scientific experiments. *SN Computer Science*, 2, 1-14.

Rucci, M., & Victor, J. D. (2015). The unsteady eye: an information-processing stage, not a bug. *Trends in Neurosciences*, 38(4), 195-206.

Schwarz, N., & Sudman, S. (Eds.). (2012). *Context effects in social and psychological research*. Springer Science & Business Media.

Shumway, R. H., Stoffer, D. S., Shumway, R. H., & Stoffer, D. S. (2017). ARIMA models. Time series analysis and its applications: with R examples, 75-163.

Streiner, D. L., Sass, D. A., Meijer, R. R., and Furman, O. (2016). The pitfalls of factor analysis. In *The palgrave handbook of child mental health*, eds M. O'Reilly and J. Lester (London: Palgrave Macmillan), 491–508.

Thiel, S. D., Bitzer, S., Nierhaus, T., Kalberlah, C., Preusser, S., Neumann, J., ... & Pleger, B. Hysteresis as an implicit prior in tactile spatial decision making. *PLoS One*, 2014, 9(2), e89802.

Tourangeau, R., Couper, M. P., & Conrad, F. (2004). Spacing, position, and order: Interpretive heuristics for visual features of survey questions. *Public Opinion Quarterly*, 68(3), 368-393.

Visser, I., Raijmakers, M. E., & Molenaar, P. (2002). Fitting hidden Markov models to psychological data. *Scientific Programming*, 10(3), 185-199.